\documentclass[twocolumn,pra,aps,showpacs]{revtex4}
\usepackage{amsmath}
\usepackage{epsfig}

\begin{document}

\title{Periodic Modulation Effect on Self-Trapping of Two weakly coupled
Bose-Einstein Condensates}
\author{Guan-Fang Wang$^{1,2}$ Li-Bin Fu$^1$ and Jie Liu$^1$}
\affiliation{$^1$Institute of Applied Physics and Computational Mathematics, P.O. Box
8009 (28), 100088 Beijing, China}
\affiliation{$^2$Institute of Physical Science and Technology, Lanzhou University, 730000
Lanzhou, China}

\begin{abstract}
With phase space analysis approach, we investigate thoroughly the
self-trapping phenomenon for two weakly coupled Bose-Einstein condensates
(BEC) in a symmetric double-well potential. We identify two kinds of
self-trapping by their different relative phase behavior. With applying a
periodic modulation on the energy bias of the system we find the occurrence
of the self-trapping can be controlled, saying, the transition parameters
can be adjusted effectively by the periodic modulation. Analytic expressions
for the dependence of the transition parameters on the modulation parameters
are derived for high and low frequency modulations. For an intermediate
frequency modulation, we find the resonance between the periodic modulation
and nonlinear Rabi oscillation dramatically affects the tunnelling dynamics
and demonstrate many novel phenomena. Finally, we study the effects of
many-body quantum fluctuation on self-trapping and discuss the possible
experimental realization of the model.
\end{abstract}
 \maketitle

\section{Introduction}

Double well system is a paradigm model used to demonstrate
marvellous quantum tunnelling property, e.g., cold atoms in
optical lattice and magnetic spins in external magnetic field
\cite{grifoni,liub}. The realization of dilute Bose degenerate gas
in 1995 provides a new opportunity to revisit the old topics
\cite{franco,anderson}. It should be addressed here, the
tunnelling of Bose-Einstein condensates (BECs) system differs from
the traditional quantum tunnelling mentioned above in two
essential aspects: system scale is macroscopic (around 10$\mu m$),
and more importantly it is a many-body system where the
interaction between atoms plays an important role. A natural
question arises that how the interaction between condensed atoms
affect the quantum dynamics in double well system. This problem
attracts much theoretical attention in past few years \cite
{milburn} and recent realization of the BECs in optical double
well trap bring new research surge \cite{shin,michael}.

Among many findings, self-trapping is most interesting one \cite%
{selftrap,raghavan,raghavan1,sigmund}. It says, the BECs atoms in
a symmetric double-well potential may shows highly asymmetric
distribution as if most atoms are trapped in one well, even under
a repulsive interaction between the degenerate atoms. Recently,
this
counter-intuition phenomenon has been successfully observed in experiment %
\cite{michael}.

In this paper, with phase space analysis approach we study this phenomenon
thoroughly. We emphasize the effects of a periodic modulation that is
applied on the energy bias of the system. We find the external periodic
field with high or low frequency can effectively modulate the transition
parameters of the self-trapping. Moreover, for an intermediate frequency
modulation, we find the BECs alternate between self-trapping regime and
non-self-trapping regime with increasing atomic interaction parameter. This
fact is attributed to chaos in phase space induced by the resonance between
the periodic modulation and nonlinear Rabi oscillation. Furthermore, we
consider the effects of many-body quantum fluctuation on self-trapping and
find the quantum fluctuation will dramatically influence the self-trapping
phenomenon. Especially for an intermediate frequency manipulation, the
alternate phenomenon does not appear since there are no chaos in
full-quantum description.

The paper is organized as follows. Self-trapping effects in two weakly
coupled BECs is introduced and analyzed with phase space approach in section
II. Section III discusses periodic modulation of this phenomenon for three
cases: high frequency, low frequency and strong resonating region,
respectively. After studying the effect of quantum fluctuation in section
IV, we give the capable observation and application of our findings in
section V. The acknowledgment is in the final part.

\section{Self-trapping in two weakly coupled BECs}

For two weakly coupled BECs, the wave function evolving with time can be
described with the superposition of individual wave function in each well.
The spacial variable and the time one of the individual wave function are
assumed to be separable. Under these approximations, we can obtain a
two-mode Schr\"{o}dinger equation, i.e., two-mode Gross-Pitaevskii equation
(GPE):

\begin{equation}
i\frac{d}{dt}\left(
\begin{array}{c}
a \\
b%
\end{array}%
\right) =H\left(
\begin{array}{c}
a \\
b%
\end{array}%
\right) ,  \label{equation1}
\end{equation}%
where $a\ $and $b\ $are the probability amplitudes of atoms in two wells
respectively. The Hamiltonian is given by%
\begin{equation}
H=\left(
\begin{array}{cc}
\frac{\gamma }{2}-\frac{c}{2}\left( \left| b\right| ^{2}-\left| a\right|
^{2}\right) & -\frac{v}{2} \\
-\frac{v}{2} & -\frac{\gamma }{2}+\frac{c}{2}\left( \left| b\right|
^{2}-\left| a\right| ^{2}\right)%
\end{array}%
\right) ,  \label{euqation2}
\end{equation}%
where $v$ is the coupling constant between the two condensates in
double-well potential, $\gamma $ is the energy bias between the two wells, $%
c $\ is the nonlinear parameter describing the interaction \cite%
{biaowu,LiBinFu,JieLiu,JieLiu0}. The total probability is conserved and set
to be $1$. Our discussion focuses on the symmetric double-well trapped BECs
system, so $\gamma =0$.

With $a=\left| a\right| e^{i\theta _{a}},b=\left| b\right| e^{i\theta _{b}}$%
, the Schr\"{o}dinger equation can be casted into a classical Hamiltonian
system by introducing the population difference $s=\left| b\right|
^{2}-\left| a\right| ^{2}$ and the relative phase $\theta =\theta
_{b}-\theta _{a}$,%
\begin{equation}
H=\gamma s-\frac{c}{2}s^{2}+v\sqrt{1-s^{2}}\cos \theta  \label{equation3}
\end{equation}%
with $s$, $\theta $ are the canonically conjugate variables of the classical
Hamiltonian system.

Self-trapping motion refers to the trajectories whose average population
difference is not zero $<s>\ne 0$. This can be well understood in phase
space of the classical Hamiltonian system (\ref{equation3}). Three kinds of
cases will emerge with different parameters, shown in Fig.1.

1) For the weak interaction, i.e., $c/v<1$, in the phase space, there are
two fixed points $p_{1}$, $p_{2}$ at $s=0$, $\theta =\pi $ and $s=0$, $%
\theta =0$, respectively (Fig.1(a)). The atoms distribution corresponding to
them is in equilibrium. The trajectories around them correspond to atoms
oscillation between the two wells, i.e., Josephson oscillation, and the
average of population difference to time is zero, i.e., $\left\langle
s\right\rangle =0$. No self-trapping effect can be found.

2) For the interaction $2>c/v>1$, two more fixed points appear on the line $%
\theta =\pi $. Among them, $p_{1}$, $p_{3}$ are stable and $p_{4}$\ is
unstable (Fig.1(b)). They are at $s=-k_{1}$, $k_{1}$, $0$, respectively,
where $k_{1}=\sqrt{1-\left( \frac{v}{c}\right) ^{2}}$. Obviously for fixed
points $p_{1}$, $p_{3}$ atoms distributions are not in equilibrium. The
average of population difference about the trajectories around them is
non-zero, i.e., $\left\langle s\right\rangle \neq 0 $. It indicates atoms
self-trapped in one well. Because the population difference $s$ and the
relative phase $\theta $ oscillate around fixed points, we denote it as
oscillating phase self-trapping.

3) With the interaction becoming stronger, i.e., $\frac{c}{v}>2$, new
trajectories emerge, for example the trajectory across point $p_{c}$
(Fig.1(c)). For these trajectories, $s$ changes with time in $\left[ -1,0%
\right] $ or in $\left[ 0,1\right] $, relative phase $\theta $
varies monotonously. Obviously $\left\langle s\right\rangle \neq
0$, atoms are self-trapped in one well, called running phase
self-trapping predicted in the reference
\cite{milburn,raghavan,MartinHolthaus} and observed in experiment
\cite{michael}.
\begin{figure}[!tbh]
\begin{center}
\rotatebox{0}{\resizebox *{8.0cm}{10.0cm} {\includegraphics
{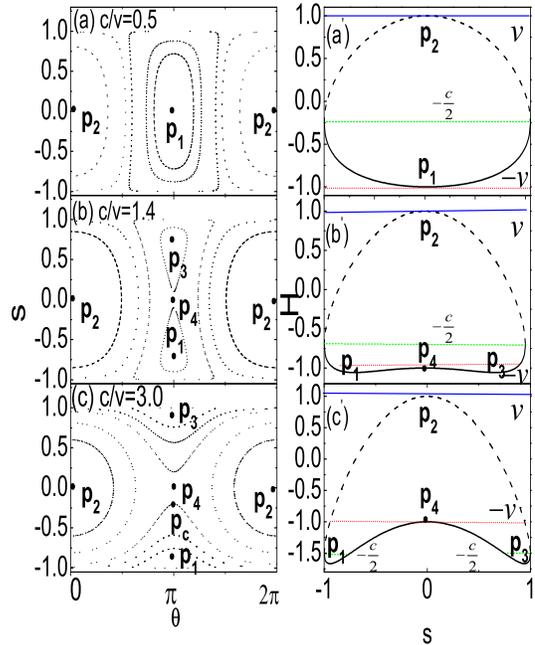}}}
\end{center}
\caption{ Trajectories on the phase space of the Hamiltonian system (3)
(left column). In right column we plot the energy profiles for the relative
phase $\protect\theta=0$ (dashed) and $\protect\theta=\protect\pi$ (solid),
respectively}
\label{fig.1}
\end{figure}

\begin{figure}[!tbh]
\begin{center}
\rotatebox{0}{\resizebox *{5.0cm}{5.0cm} {\includegraphics
{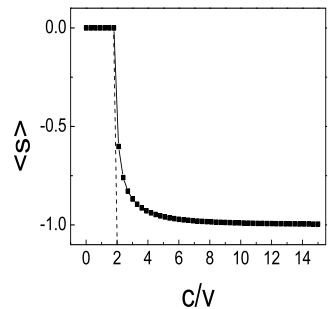}}}
\end{center}
\caption{The average of population difference vies the parameter
$c/v$ when the trajectory $s=-1,\protect\theta =0$ evolves with
time. } \label{fig2}
\end{figure}

The above changes on the topological structure of the phase space associates
with the change of the profile of the physical energy. With considering the
relative phase being zero or $\pi $, the energy depending on the population
difference $s$ can be derived from Eq.(3) and plotted in the right column of
Fig.1 for varied parameters. The transition from case 1) to case 2)
corresponds to the bifurcation of the energy profile of $\theta =\pi $: The
energy function of single minimum bifurcates into the function of two
minimums; The low limit of the energy profile with $\theta =0$ is $-c/2$,
and the energy of the saddle point $P_{4}$ is $-v$, which is the local
maximum of the energy profile with $\theta =\pi $. The transition from case
2) to case 3) is characterized by the overlap of the energy regime of the
two profiles. In this case the trajectory originated from $s=-1,\theta =0$
can not cross the energy hill peaked by the saddle point. It will be
confined in the low phase plane, corresponding the ''running phase'' type
self-trapping.

From the above analysis on the energy function we can obtain a general
criteria for the self-trapping trajectories, that is $H(s,\theta )<-v$. With
it, the transition parameters of self-trapping for the trajectory with
arbitrary initial value can be found. For example, in the situation of
recent experiment \cite{michael} $s=0.5,\theta =0$, the $c/v=15.0$ is the
critical point for the transition to the 'running phase' type self-trapping.
For the state of initial condition $s=-1,\theta =0$ the transition parameter
is $c/v=2.0$ in our analysis, agrees with our numerical simulations as shown
in Fig.2. The scaling law near the transition point shows a logarithmic
singularity.

\section{Periodic modulation of self-trapping}

Generally, we can control the behavior of the system by applying a periodic
modulation. For example tunnelling probability can be controlled through
adjusting the modulation parameters for the linear case \cite%
{Grossman,Gomez,Yosuke,duan}. In this section, we will discuss how a
periodic modulation affect the nonlinear self-trapping. Without losing
generality, we assume that the modulation is applied on the energy bias with
amplitude $A$ and frequency $\omega$, i.e., $\gamma=A\sin\omega t$. Then the
nonlinear GPE still can be mapped into the time dependent Hamiltonian (3).
We then can investigate the global property of the trajectories with the
phase space analysis method as in above section. Differently, in this case,
the phase space of the time-dependent Hamiltonian (or Poincare section) is
drawn by stroboscopic plotting of the trajectories at the moment of integer
multiple times of the modulation period, i.e., $2\pi/\omega$. We will
consider following three typical cases corresponding to different modulation
frequency regimes.

\subsection{High frequency modulation $(\protect\omega \gg v)$}

When the system is added by a high frequency modulation, the two kinds of
self-trapping, i.e., oscillating or running phase, still exist as shown in
Fig.3. Here the phase space are stroboscopic plotting of the trajectories at
the moment of integer multiple times of the modulation period. Compared with
Fig.1, the fixed points are shifted to the leftside. This is because the
phase space depends on the time moment when Poincare surface of section is
made. For the different time moment, the topological structure of the
Poincare section keeps consistent which guarantee the validity of our
observation. Moreover, we find that the critical interaction parameter above
which the self-trapping occurs changes dramatically with periodic
modulation. Fig.4(a), taking the trajectory $s=-1,\theta =0$ as an example,
shows the critical interaction is about 1.531 at $\omega =100,A/\omega
=1.0,v=1.0$. Fig.4(b) shows phase diagram for the transition to the
self-trapping. It says that the critical values $C/V$ depend on the
modulation parameters in the way of the  Bessel function $J_{0}(\frac{A}{%
\omega })$.

\begin{figure}[!tbh]
\begin{center}
\rotatebox{0}{\resizebox *{5.0cm}{7.0cm} {\includegraphics
{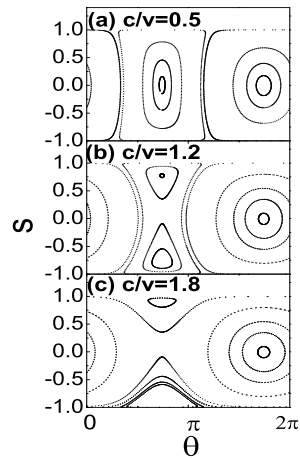}}}
\end{center}
\caption{Phase space of the Hamiltonian system (3) with $\protect\gamma=Asin%
\protect\omega t$ at $\protect\omega =100,A/\protect\omega =1.0$, obtained
by stroboscopic plotting the trajectories with period $2\protect\pi/\protect%
\omega$.} \label{fig3}
\end{figure}

\begin{figure}[!tbh]
\begin{center}
\rotatebox{0}{\resizebox *{8.0cm}{5.5cm} {\includegraphics
{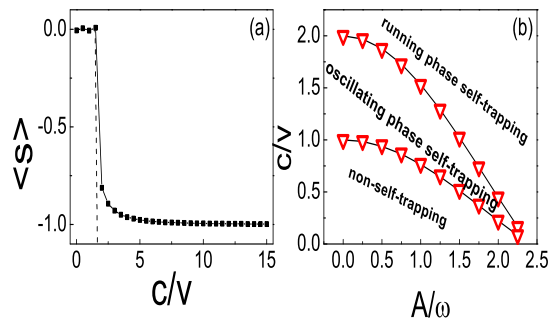}}}
\end{center}
\caption{(a) The average of population difference vies parameter $c/v$ when
the trajectory $s=-1,\protect\theta =0$ evolves with time at $\protect\omega %
=100,A/\protect\omega =1.0$. (b) Transition phase diagram for the occurrence
of the self-trapping. The triangle describes the case at $\protect\omega =100
$. The line describes $J_{0}\left( A/\protect\omega \right) $ (lower line)
and $2J_{0}\left( A/\protect\omega \right) $ (upper one). The situation at $%
\protect\omega =10,1000$ also agree with the theoretical
prediction through our calculation. } \label{fig4}
\end{figure}

The above observation can be understood from following deduction. Using high
frequency approximation, the time-dependent system (\ref{equation1}) is
equivalent to a stationary one. Let us make the transformation $a=e^{i\frac{A%
}{2\omega }\cos \omega t}a^{^{\prime }}$, $b=e^{-i\frac{A}{2\omega }\cos
\omega t}b^{^{\prime }}$. Then we can obtain the following Schr\"{o}dinger
equation

\begin{equation}
i\frac{da^{^{\prime }}}{dt}=-\frac{c}{2}(\left| b^{^{\prime }}\right|
^{2}-\left| a^{^{\prime }}\right| ^{2})a^{^{\prime }}-\frac{v}{2}e^{i\frac{A%
}{\omega }\cos \omega t}b^{^{\prime }},  \label{euqation5}
\end{equation}
\begin{equation}
i\frac{db^{^{\prime }}}{dt}=\frac{c}{2}(\left| b^{^{\prime }}\right|
^{2}-\left| a^{^{\prime }}\right| ^{2})b^{^{\prime }}-\frac{v}{2}e^{i\frac{A%
}{\omega }\cos \omega t}a^{^{\prime }}.  \label{equation6}
\end{equation}%
Using formula
\begin{equation}
e^{\pm iz\cos \omega t}=\sum_{n=-\infty }^{\infty }J_{n}\left( z\right)
\left( \pm i\right) ^{n}e^{\pm in\omega t},  \label{equation7}
\end{equation}%
where $J_{n}\left( z\right) $ is the Bessel function of nth order, and
considering the contribution of those higher order Bessel function to the
integrals can be neglected \cite{Yosuke1,Yosuke}, the effective coupling
constant becomes $v^{^{\prime }}=vJ_{0}(\frac{A}{\omega })$.

Immediately, we can obtain the critical values are $\frac{c}{v}=J_{0}\left(
\frac{A}{\omega }\right) $ for oscillating phase self-trapping and $\frac{c}{%
v}=2J_{0}\left( \frac{A}{\omega }\right) $ for running phase self-trapping
based on the results in section II. This result is consistent with the
numerical calculation.

\subsection{Low frequency modulation $\left( \protect\omega \ll v\right)$}

For this case, it is not significant to identify the two kinds of
self-trapping. We suppose initially all particles confined in one
well with the relative phase of $\pi$, i.e., $s=-1,\theta=\pi$.
Then for given parameters we check if the trajectory is self
trapped, $<s> \ne 0$. We find that the self-trapping still occurs
for some regime of parameters, as shown in Fig.5(a). There, we
see, above a critical value of interaction which is a function of
modulation amplitude and the energy gap, the average population
difference jump from a irregular oscillation to a constant
($\left\langle s\right\rangle =-1$, self-trapping regime).

To see what happened before and after the transition, we plot the
phase space of the system with $\gamma$ changing slowly, seeing
Fig.6. Below the critical point, i.e., $c/v=13.5$, the fixed
$p_{1}$ (where we started from) moves up
smoothly, at a certain $\gamma_c $, it collides with another fixed point $%
p_{4}$ and gives birth to a new trajectory. Above the critical point, i.e., $%
c/v=13.6$, the trajectory $p_{1}$ moves along the line $\theta=\pi$ up and
down smoothly, having no chance to collide with other fixed point. For the
large $c$ case, the shift of our trajectory away from the bottom line ($s=-1$%
) is small, so that the time average of the population difference is
approximately $-1$.

The singularity point $\gamma_c$ where two fixed points collide
leading to the topological change on phase space depends on the
modulation amplitude but little on its frequency. Our theoretical
deduction gives $\gamma =\pm \left( c^{2/3}-v^{2/3}\right)
^{3/2}$\cite{JieLiu}. When the modulation frequency
is very small, the parameter $\gamma $ varies adiabatically with time $\dot{%
\gamma}=A\omega \cos \omega t\ll v$. If amplitude $A$ does not exceed the
critical values, atoms should be self-trapped. This gives the condition $%
c>\left( A^{2/3}+v^{2/3}\right) ^{3/2}$ for self-trapping, which
is well coincident to the numerical result shown in Fig.5(b).
Below the critical point, the collision between the two fixed
points mean occurrence of the instability, which may leads to the
irregular oscillation of the average population difference as seen
in Fig.5(a).

\begin{figure}[!tbh]
\begin{center}
\rotatebox{0}{\resizebox *{8.0cm}{5.5cm} {\includegraphics
{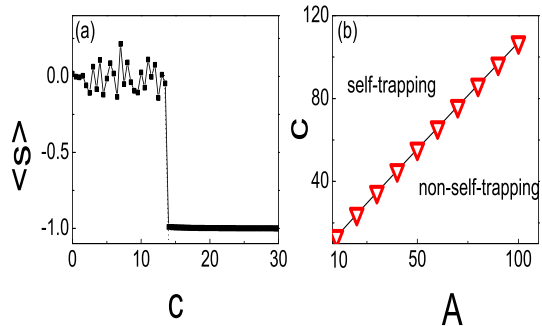}}}
\end{center}
\caption{(a) The average of population difference vies the
interaction parameter at $\protect\omega =0.01,A=10,v=1.0$. (b)
The relationship between the critical values of $c$ and $A$ when
$\protect\omega $ is extremely small at $v=1.0$. The triangle
describes the situation at $\protect\omega =0.01$. The line
describes the theoretical prediction, $c=\left(
A^{2/3}+v^{2/3}\right) ^{3/2}$.} \label{fig5}
\end{figure}

\begin{figure}[!tbh]
\begin{center}
\rotatebox{0}{\resizebox *{8.0cm}{10.0cm} {\includegraphics
{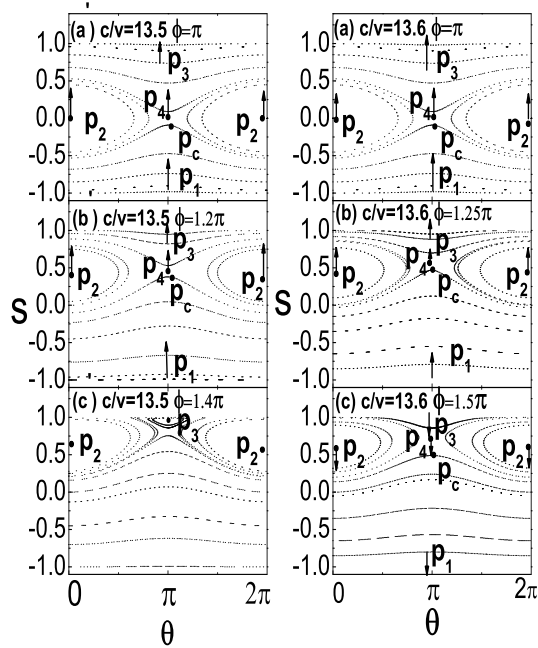}}}
\end{center}
\caption{Evolution of the phase space motions of the Hamiltonian
system (3) with $\protect\gamma =A\sin \protect\omega t$ at
$A=10,c/v=13.5$ (left collum) and $A=10,c/v=13.6$ (right collum)
as $\protect\gamma $ changes adiabatically. The arrows refer to
the moving direction of the fixed points as $\protect\phi $
increases, i.e., $\protect\gamma $ changes adiabatically, where
$\protect\phi =\protect\omega t$.} \label{fig6}
\end{figure}

\subsection{Strong resonating region $\left( \protect\omega \approx v\right)
$}

For the strong resonating region, the situation is complicated.
Due to the strong resonance between the nonlinear Rabi oscillation
and the periodic modulation, chaos appears as shown in the phase
space plotting Fig.7, where the scattered points denote the
chaotic trajectories forming chaotic sea\cite{chaos}. Inside the
chaotic sea there are stable islands corresponding to the
self-trapping trajectory. We find with increasing the interaction
parameter,
the stable islands appear alternately. Fig.8 describes this alternate with $%
sign=1$ standing for self-trapping and $sign=0$ for non-self-trapping when
interaction parameter varies continuously. The mutations are the transition
parameters. Analytic expression for the dependence of the transition
parameters on the modulation parameters can not be found.
\begin{figure}[!tbh]
\begin{center}
\rotatebox{0}{\resizebox *{5.0cm}{8.0cm} {\includegraphics
{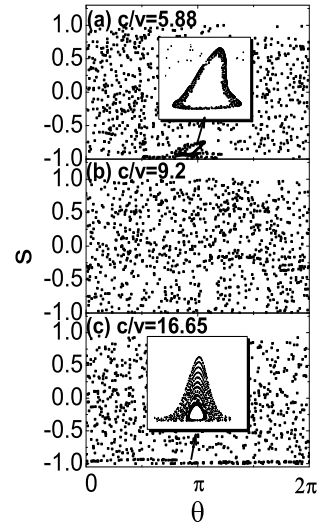}}}
\end{center}
\caption{Phase space of the Hamiltonian system (3) with $\protect\gamma %
=A\sin \protect\omega t$ at $\protect\omega =2.0$, obtained by
stroboscopic plotting the trajectories with period $2\protect\pi
/\protect\omega $. (a) describes the situation at $c=5.88$, (b) at
$c=9.2$, and (c) at $c=16.65$ respectively. The small fig is the
magnified portion in the proper big Fig..} \label{fig7}
\end{figure}

\begin{figure}[!tbh]
\begin{center}
\rotatebox{0}{\resizebox *{8.0cm}{4.5cm} {\includegraphics
{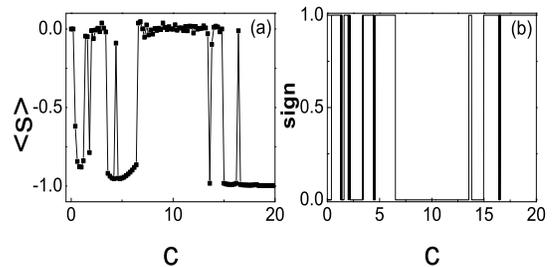}}}
\end{center}
\caption{(a) The average of population difference vies the interaction
parameter at $\protect\omega =2.0,A=10,v=1.0$. (b) Critical values of $c$
for occurrence of self-trapping shows complicated behavior at $v=1.0,\protect%
\omega =2.0,A=10$. See the text for detailed explanation.}
\label{fig8}
\end{figure}

\section{The effect of quantum fluctuation}

In the above discussions on the self-trapping, our framework is
the mean field treatment with assuming that the number of particle
is large enough. However, in practical experiment, the particle
number is finite, we want to know how the quantum fluctuation
caused by finite number of atoms affects the self-trapping. With
this motivation we investigate following second quantized
Hamiltonian which is the quantum counterpart of our mean field
system (1),
\begin{equation}
H=-\frac{\gamma }{2}\left( a^{\dagger }a-b^{\dagger }b\right) -\frac{c}{2N}%
\left( a^{\dagger }a^{\dagger }aa+b^{\dagger }b^{\dagger }bb\right) +\frac{v%
}{2}\left( a^{\dagger }b+b^{\dagger }a\right)   \label{equation11}
\end{equation}%
where $\gamma $,$c$,$v$ are energy bias between the two modes, the
interaction between atoms and the coupling between the modes, respectively. $%
a^{\dagger }$, $b^{\dagger }$ ($a$, $b$) are creating (annihilating) boson
operators. $N$ \ is the total number of atoms. Its matrix elements are
obtained in the representation of Fock states $\left| n,N-n\right\rangle $. $%
H_{n,n}=\left\langle N-n,n\right| H\left| n,N-n\right\rangle
,H_{n,n-1}=H_{n-1,n}=\left\langle N-n,n\right| H\left|
n-1,N-n+1\right\rangle $.  Corresponding to the mean field $s=-1$ our
initial state in the full quantum framework is $\left| 0,N\right\rangle $.
We then can solve the corresponding Schr\"{o}dinger equation $i\frac{d}{dt}%
a_{n}=H_{n,n-1}a_{n-1}+H_{n,n}a_{n}+H_{n,n+1}a_{n+1}$ with
Runge-Kutta method, and trace the time evolution of population
difference $s=\sum \frac{\left| a_{n}\right| ^{2}\left(
N-2n\right) }{N}$, where $a_{n}$ is the amplitude of Fock state
$|n,N-n>$.  Our main results are that, for the cases of high or
low frequency modulation, the self-trapping effect can still be
observed in full-quantum description, i.e., with increasing the
interaction parameter, the quantum average population will
transition from symmetric distribution to asymmetric distribution.
However, different from the mean field case, the transition here
is smooth or continuous, no scaling law or singularity is
observed. For the case of intermediate modulation, because the
quantum effect dramatically suppress the classical chaos, the
alternate phenomenon observed in the mean field case vanishes.
What we observe is that a continuous  transition to the
self-trapping regime with increasing the interaction parameter. In
following we give detailed presentation of our results.

For $\omega \gg v$, with increasing the interaction parameter,
Fig.9 compares the transition to self-trapping in full-quantum
treatment with that  in  mean field framework, by  tracing the
time evolution of population difference. It is found that, within
mean field framework (Fig.9(a)), there is an abrupt  transition to
self-trapping characterized by a logarithmic scaling law as
mentioned above.  However, for  full-quantum calculations (
Fig.9(b)) the transition become smooth and no scaling law or
singularity is found in this case.

\begin{figure}[!tbh]
\begin{center}
\rotatebox{0}{\resizebox *{7.0cm}{6.0cm} {\includegraphics
{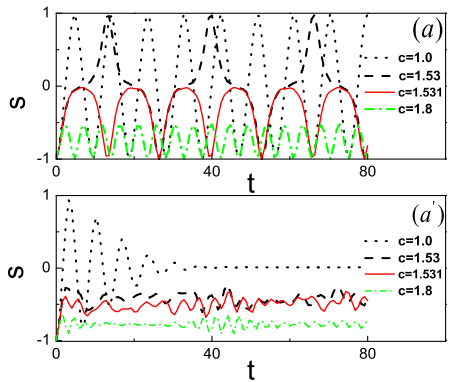}}}
\end{center}
\caption{(Color online.) Evolution of population difference with time at $%
\protect\omega =100,A/\protect\omega =1.0,v=1.0$. (a) is
mean-field results. (a$^{^{\prime }}$) is full-quantum results.}
\label{fig9}
\end{figure}

For case of $\omega \ll v$, our results are similar. Seeing
Fig.10, we find the transition to self-trapping regime is abrupt
in mean field treatment(Fig.10(a)), but is smooth for the
full-quantum case (Fig.10(b)).

\begin{figure}[!tbh]
\begin{center}
\rotatebox{0}{\resizebox *{7.0cm}{6.0cm} {\includegraphics
{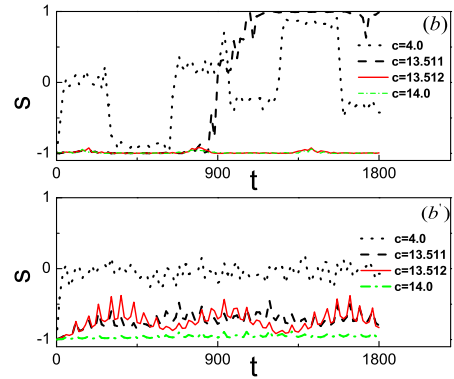}}}
\end{center}
\caption{(Color online.) Evolution of population difference with time at $%
\protect\omega =0.01,A=10,v=1.0$. (b) is mean-field results.
(b$^{^{\prime }} $) is full-quantum results. In (a) the amplitude
of line $c=14.0$ is so small that it almost can not be seen.}
\label{fig10}
\end{figure}

For the intermediate case of $\omega \approx v$, the situation is
quite
 different.  Classical chaos shown in the mean field treatment is
 expected to be suppressed by the quantum fluctuation, alternate phenomenon observed in the mean field case
 vanishes, as we see in Fig.11. In mean field method (Fig.11(a)), self-trapping
effect, $\left\langle s\right\rangle \neq 0$, and
non-self-trapping effect, $\left\langle s\right\rangle =0$\,
appear alternately with increasing the interaction parameter,
while, in full-quantum method (Fig.11(b)), a continuous transition
to the self-trapping regime is observed, that is, with increasing
the interaction parameter, the BEC distribution become more and
more self-trapped in a well.

\begin{figure}[!tbh]
\begin{center}
\rotatebox{0}{\resizebox *{7.0cm}{6.0cm} {\includegraphics
{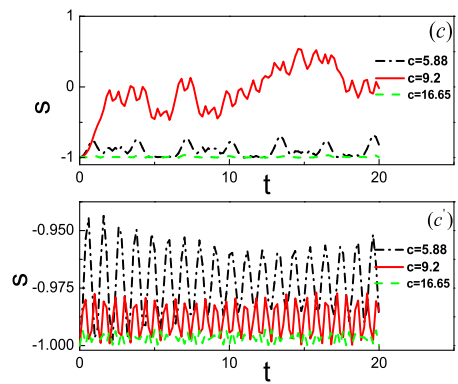}}}
\end{center}
\caption{(Color online.) Evolution of population difference with time at $%
\protect\omega =2.0,A=10,v=1.0$. (c) is mean-field results. (c$^{^{\prime }}$%
) is full-quantum results.} \label{fig11}
\end{figure}

From the above analysis, we see in practical experiment to observe
the dramatic effects of the quantum fluctuation  caused by the
finite number of particle, we need choose the parameters in the
regime of the occurrence of  classical chaos in mean field
systems.

\section{Conclusions and Discussions}

We have investigated the periodic modulation effect on the self-trapping of
two weakly coupled Bose-Einstein Condensates in a symmetric double-well
potential. Analytic expressions for the dependence of the transition
parameters on the modulation parameters are derived for high and low
frequency modulations. It is shown that the self-trapping phenomenon can be
controlled, saying, the transition parameters can be adjusted effectively by
the periodic modulation. We also study the effects of many-body quantum
fluctuation on self-trapping due to finite number of the particle. We find
that the quantum fluctuation make the critical phenomenon hazy for high and
low frequency manipulation. For an intermediate frequency, the alternative
phenomenon does not appear since there are no chaos in full-quantum
description.

Experimentally, BEC in optical double well system can be generated
by coherently splitting BEC into a double-well potential generated
by two laser beams. The radial separation of the potential wells
is controlled by the frequency difference. The energy offset can
be
periodically modulated by adjusting the intensity of the two laser beams\cite%
{michael,shin}. The second possible physical system can be used to realize
our model is the tunnelling between the internal state of BEC like $^{87}$Rb%
\cite{cornell}. There, two internal states are separated by the relatively
large hyperfine energy, but in presence of a near-resonant coupling field
the states appear to be nearly degenerate. The energy bias can be adjusted
by the detuning of the lasers from resonance. Our theory predicts the
periodic modulation on the energy bias will dramatically affect the
tunnelling dynamics and the nonlinear self-trapping. We hope our discussion
will stimulate the experiments in the direction.

\section{Acknowledgments}

This work was supported by National Natural Science Foundation of China
(Grant No.: 10474008,10445005) and by Science and Technology fund of CAEP.
We sincerely thank professors Biao Wu, Bing-Bing Wang, Pan-Ming Fu and
Shi-Gang Chen for their valuable discussions.

\end{document}